\documentclass{llncs}
\usepackage{graphicx}
\usepackage{filecontents}
\usepackage{algorithm}
\usepackage{algpseudocode}
\usepackage[compatibility=false]{caption}
\usepackage{subcaption}
\usepackage{hyperref}
\usepackage{listings}
\usepackage{color}
\usepackage{multirow}
\usepackage{booktabs}
\begin{document}%
\title{Accelerating Direction-Optimized Breadth First Search on Hybrid Architectures}%
\author{Scott Sallinen \and Abdullah Gharaibeh \and Matei Ripeanu}%
\institute{University of British Columbia \\
\email {\{scotts, abdullah, matei\}@ece.ubc.ca}}%
\maketitle%
\begin{abstract}%
Large scale-free graphs are famously difficult to process efficiently: the skewed vertex degree distribution makes it difficult to obtain balanced partitioning. Our research instead aims to turn this into an advantage by partitioning the workload to match the strength of the individual computing elements in a Hybrid, GPU-accelerated architecture. As a proof of concept we focus on the direction-optimized breadth first search algorithm. We present the key graph partitioning, workload allocation, and communication strategies required for massive concurrency and good overall performance. We show that exploiting specialization enables gains as high as 2.4x in terms of time-to-solution and 2.0x in terms of energy efficiency by adding 2 GPUs to a 2 CPU-only baseline, for synthetic graphs with up to 16 Billion undirected edges as well as for large real-world graphs. We also show that, for a capped energy envelope, it is more efficient to add a GPU than an additional CPU.  Finally, our performance would place us at the top of today's [Green]Graph500 challenges for Scale29 graphs.  
\end{abstract}%
\section{Introduction}%
\label{sec:intro}%
Large-scale graph processing plays an important role for a wide range of applications where internal data structures are represented as graphs: from social networks, to protein-protein interactions, to the neuron structure of the human brain. A large class of real-world graphs, the \textit{scale-free graphs}, have a heterogeneous vertex degree distribution: most vertices have a low degree but a few vertices are highly connected \cite{wang2003complex} \cite{jeong2001lethality} \cite{bullmore2009complex}.

Breadth First Search (BFS) is the building block for many graph algorithms (e.g., Betweenness Centrality,  Connected Components) and it has similar structural properties to other algorithms (e.g., Single Source Shortest Paths - SSSP). BFS also exposes the main challenges in graph processing: data-dependent memory accesses, low compute-to-memory access ratio, and low memory access locality. Additionally, for scale-free graphs, the amount of parallelism exposed is highly heterogeneous and data-dependent. For these reasons the Graph500 \cite{Graph500} and GreenGraph500 \cite{GreenGraph500} graph processing competitions have adopted BFS on scale-free graphs as their main benchmark to compare the efficiency of graph processing platforms in terms of time-to-solution and energy.

Recently, Beamer et al. \cite{beamer2011searching} introduced the \textit{direction-optimized} BFS algorithm that takes advantage of the scale-free property (Section \ref{sec:dirop}). This algorithm combines the classic \textit{top-down} BFS traversal, with inverse \textit{bottom-up} steps and offers a sizable speedup. To date, however, all implementations have focused either on CPU-only platforms \cite{yasui2014fast} or require that the graph fits \textit{entirely} in the accelerator memory \cite{you2014designing}.

Our past work \cite{gharaibeh2014efficient} tests the hypothesis that assembling processing elements with diverse characteristics (i.e., massively-parallel processors optimized for high throughput, and traditional processors optimized for fast sequential processing) is a good match for scale-free graph workloads. While we have proven this hypothesis for a wide set of algorithms (including traditional top-down BFS, Connected Components, SSSP and PageRank), direction-optimized BFS poses new challenges: (i) as it is up to one order of magnitude faster than traditional BFS, it stresses the communication channels between the processing elements of the heterogeneous platform, exposing new bottlenecks; (ii) it requires both pull and push access to vertex state that has to be efficiently exposed by the supporting middleware; (iii) as the processing elements do not share memory, a low-overhead solution must be designed to coordinate them to switch between bottom-up and top-down phases of the algorithm, and finally, (iv) it requires specialized graph partitioning and workload allocation strategies that match the characteristics of the workload to those of the processing elements.

This paper makes the following contributions:%
\begin{enumerate}%
\setlength{\itemsep}{0pt}%
\setlength{\parskip}{0pt}%
\item It provides further evidence that specialization -- i.e., intelligent graph partitioning such that the resulting workload matches a heterogeneous set of processing elements -- is key to extracting maximum efficiency when facing a fixed cost or energy constraint. (Sections \ref{sec:partitioning}, \ref{sec:hardwareconf})%
\item It extends \textsc{Totem}, our heterogeneous graph processing engine, to support a new class of frontier-based algorithms which require exposing both push and pull access to distributed vertex state. (Section \ref{sec:algorithms})
\item It introduces optimizations key to boost the performance of direction-optimized BFS on a heterogeneous platform: partitioning and workload allocation, communication reduction, and improving access locality. (Sections \ref{sec:partitioning}, \ref{sec:swapping}, \ref{sec:reindexing})
\end{enumerate}%
We evaluate these techniques across multiple hardware configurations and multiple large-scale graph workloads. Our evaluation shows an improvement of time-to-solution by up to 2.4x and energy efficiency by up to 2.0x against a CPU-only implementation, and compares favorably against state-of-the-art single node solutions (e.g., \textsc{Galois}). (Sections \ref{sec:realworld}, \ref{sec:energy})%
\section{Challenges and Opportunities}%
\label{sec:opportunities}%
The key difficulty when processing scale-free graphs is a result of the heterogeneous vertex connectivity. (For example, over 70\% of all vertices in the Twitter follower graph \cite{cha2010measuring} have less than 40 in/out edges. The remaining vertices have increasingly large connectivity: the largest having over 3 million edges.) This property makes obtaining balanced partitions difficult, as generally the memory footprint of a partition is proportional with the number of edges allocated to it, while the processing time is a complex function that depends on the number of vertices and edges allocated to the partition, and the specific properties of the workload (e.g., compute intensity, access locality).

Past work has generally assumed a homogeneous compute platform and has prioritized balancing partitions in terms of size \cite{pearce2013scaling}. This leads, however, to unbalanced partitions in terms of processing time due to the high-connectivity nodes. Recent strategies such as 'high degree vertex delegation' \cite{pearce2014faster} continue to assume a homogeneous platform and aim for better load-balancing while dealing with high degree vertices.
\subsection{Improving Performance with Hardware Accelerators}%
\label{sec:harwareaccelerate}%
A GPU-accelerated system offers the opportunity to benefit from heterogeneity: instead of attempting to balance partitions by evenly distributing the workload based on memory footprint, one can choose to 'embrace' heterogeneity and partition such that the workload generated by a partition matches best the strengths of the processing element the partition is allocated to -- e.g., by creating partitions that expose massive parallelism and allocating them to a GPU \cite{gharaibeh2014efficient} \cite{cumming2014application}.

However, efficiently harnessing a GPU-accelerated setup brings new challenges: First, it is difficult to design partition and allocation strategies that harness the platform efficiently. Second, GPUs tend to have over an order of magnitude less memory than the host and cannot process large partitions. (A key constraint -- for example, the edge list of a Scale30 graph, a synthetic graph used in the Graph500 benchmark, occupies 256GB in the memory-efficient Compressed Sparse Row format, yet a Kepler K40 GPU has only 12GB of memory).

Note that past projects have explored GPU solutions, but either assume that the graph fits the memory of one \cite{hong2011accelerating} \cite{you2014designing}, or multiple GPUs \cite{merrill2012scalable}. In both cases, due to the limited memory space available, the scale of the graphs that can be processed is significantly smaller than the large graphs presented in this paper.

In any case, techniques that aim to optimize graph processing on the GPU are complementary to the approach proposed in this work in that they can be applied to the compute kernels to improve the overall performance of the hybrid system. In fact, this work uses the "virtual warp" technique proposed by Hong et al. \cite{hong2011accelerating} which aims to reduce divergence among the threads of a warp and hence improve the GPU kernel's performance.
\subsection{Improving Performance with Direction-Optimized BFS}%
\label{sec:dirop}%
Similar to other graph algorithms, level-synchronous Breadth First Search (BFS) exposes the concept of a \textit{frontier}: the set of active vertices, which, for BFS, form the current level. To discover the next level, i.e., the next frontier, the traditional top-down BFS approach explores all edges of the vertices in the  current frontier and builds the next frontier as the new vertices that can be reached (i.e., the vertices that have not been visited before). For scale-free graphs, this can cause high write traffic as many edges out of the current frontier can attempt to add the same vertex into the next frontier.

\textit{Direction-optimized} BFS \cite{beamer2011searching} is based on the key observation that the next frontier can also be built in a different, \textit{bottom-up} way: by iterating over the vertices that have never been activated and selecting those that have a neighbour in the current frontier. Depending on graph topology and the current state of the algorithm, a bottom-up step can improve performance for two reasons. First, it can result in exploring fewer edges: once it has been determined that a vertex has a neighbour in the frontier it is not necessary to visit its other edges, thus reducing work particularly for high degree vertices. Second, it generates less contention as the thread that updates a vertex state (i.e., marks it as belonging to the new frontier) only reads its neighbour's state but does not update it.
\setlength{\textfloatsep}{-5pt}
\begin{figure}%
	\centering%
	\begin{subfigure}{.5\textwidth}%
		\centering%
		\includegraphics[width=0.98\linewidth,trim=4 2 1 3,clip]{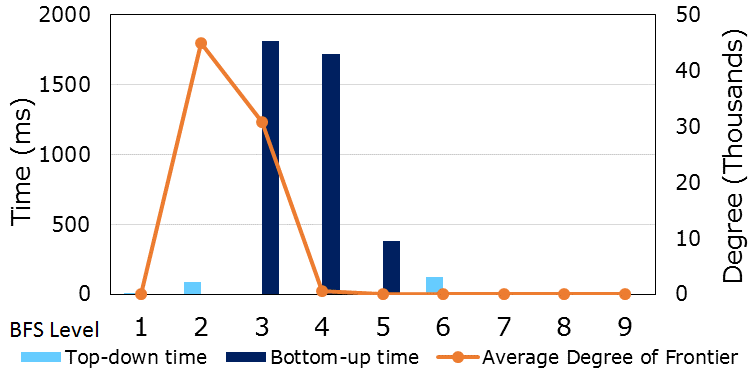}%
		\label{fig:frontierscale30}%
	\end{subfigure}%
	\begin{subfigure}{.5\textwidth}%
		\centering%
		\includegraphics[width=0.98\linewidth,trim=4 2 4 3,clip]{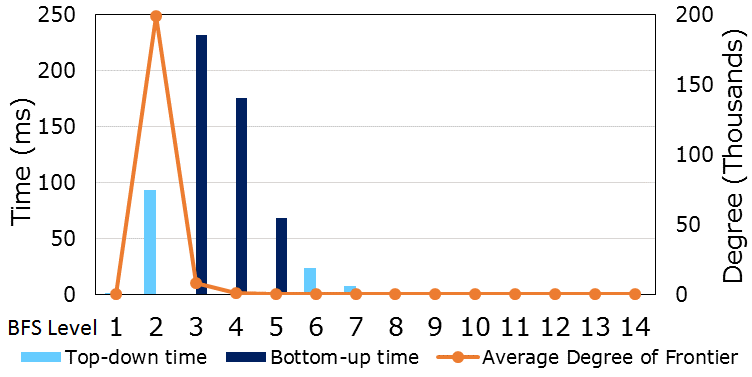}%
		\label{fig:frontiertwitter}%
	\end{subfigure}%
	\caption{Processing time per BFS level (left axis) and the average degree of vertices in the frontier (right axis). Graphs: \textit{Left} Synthetic Scale30. \textit{Right} Twitter \cite{cha2010measuring}.}%
	\label{fig:frontier}%
\end{figure}%

During BFS processing of a scale-free graph, the vertices with high connectivity are quickly reached. Next, with these \textit{few high degree} vertices in the frontier, the number of vertices in the next frontier will be large. At this point, it becomes more efficient to employ a bottom-up step: this will reduce the number of edges explored, and will alleviate write pressure corresponding to the many vertices that will be added to the frontier. Figure \ref{fig:frontier} supports this observation: the average degree of the frontier is large immediately after start (i.e., during the initial top-down steps of direction-optimized BFS). Next, during the bulk of the computation time, the average degree of a processed vertex is lower and continues to decrease; as a result, bottom-up steps are more efficient, but in effect, the many low degree vertices may end up being processed unsuccessfully at multiple BFS levels before they are finally included in the frontier. As the average degree of the frontier continues to decrease, top-down processing becomes again more efficient during the last few steps of the algorithm.
\section{The Hybrid Algorithm}%
\label{sec:algorithms}%
To harness the opportunities offered by a heterogeneous platform, several issues need to be addressed: (i) partitioning the graph and allocating partitions to processing elements to match their strengths and limits (i.e., massive parallelism yet limited accelerator memory); (ii) efficiently communicating between partitions; and (iii) coordinating the participating processing elements. 

Our past work \cite{gharaibeh2014efficient} demonstrates that the benefits of using a heterogeneous platform exceed the cost of communication between partitions hosted on different processing elements: the reduction in compute time obtained by adding GPUs is far larger than the added communication time over the PCI bus to synchronize between partitions. However, direction-optimized BFS adds new challenges: first, the bottom-up steps are up to one order of magnitude faster than the equivalent top-down steps thus potentially expose the communication over the PCI bus as the key bottleneck. Second, all processing elements participating in a direction-optimized BFS computation need to coordinate and chose direction at the same time. This coordination can add further communication overheads, as there is no shared state between the processing elements. 
\subsection{Direction-optimized BFS for Partitioned Graphs}%
\label{sec:dobfs}%
Our BFS algorithm \ref{alg:bfs} is based on the Bulk Synchronous Parallel (BSP) model which supports well a system setup where processing elements that do not share memory, and where the graph needs to be partitioned for processing. Each BFS level of the algorithm contains a communication operation: the top-down steps use a Push-based method; the bottom-up steps a Pull-based method (described in algorithms \ref{alg:scatter} and \ref{alg:gather}). The partitions each have an array of frontiers corresponding to other partitions. In addition, vertices have an associated partition ID, allowing the algorithm to determine whether or not the vertex is remote, and which frontier to use.

\textit{Top-down steps} explore the edges of the vertices in the frontier of the local partition: each such vertex activates -- mark as belonging to the next frontier -- a vertex that is either local or remote (belonging to another partition). During the \textit{push} phase, the remote information is passed to the corresponding partitions, then all partitions wait for synchronization before starting the next round to ensure they all have pushed all necessary information.

\textit{Bottom-up steps} start by aggregating, in each partition, the global frontier by \textit{pulling} the required information from all other partitions. Then each partition completes its level by checking its local not-yet-activated vertices against this global frontier. Then partitions synchronize: this ensures that all partitions have completed creating their new local frontier and are ready for the next round.

\textit{Optimizations.} Key performance gains are achieved due to batch communication and message reduction: the push and pull operations only happen once per BSP round, and only to remote neighbours (i.e., only the data relevant to remote partitions). An additional optimization we apply is specific to the case when the user requires computing the BFS traversal tree as in the Graph500 benchmark (and not only labeling nodes with their 'depth'): To reduce communication overhead, the parent of a vertex is not communicated during the traversal but is collected from the different address spaces in a final aggregation step (only the visited status is updated during traversal).\\

\begin{minipage}{\textwidth}
\captionof{algorithm}{Direction-optimized BFS algorithm for partitioned graphs.}
\label{alg:bfs}
\begin{minipage}{0.60\textwidth}
\begin{lstlisting}
func BFS_Kernel (Partition PID, StepType STEP)
 if (STEP == TOP-DOWN) then
   parallel foreach (Vertex in Frontier[PID]) do
   @$|$@ foreach (Nbr in Vertex.Neighbours) do
   @$|$@ @$|$@ if (!Nbr.isVisited) then
   @$|$@ @$|$@   NextFrontier[Nbr.partition].Add(Nbr)
   @$|$@ @$|$@   BFSParentTree[Nbr.partition][Nbr] = Vertex
   @$|$@ @$|$@   Nbr.isVisited = true
   @$|$@ @$|$@ end if
   @$|$@ end for
   end parallel for
   PushFrontiers(PID)
 else if (STEP == BOTTOM-UP) 
   PullFrontiers(PID)
   parallel foreach (Vertex in Partition[PID]) do  @\label{alg:bfs:line:allvertices}@
   @$|$@ if (!Vertex.isVisited) then
   @$|$@   foreach (Nbr in Vertex.Neighbours) do
   @$|$@   @$|$@ if (Vertex in Frontier[Nbr.partition]) then
   @$|$@   @$|$@   NextFrontier[PID].Add(Vertex)
   @$|$@   @$|$@   BFSParentTree[PID][Vertex] = Nbr
   @$|$@   @$|$@   Vertex.isVisited = true
   @$|$@   @$|$@   break for
   @$|$@   @$|$@ end if
   @$|$@   end for
   @$|$@ end if
   end parallel for
 end if
 Synchronize()
end func
\end{lstlisting}
\end{minipage}
\begin{minipage}{.39\textwidth}
 \newline
\captionof{algorithm}{Push Frontiers}
\label{alg:scatter}
\begin{lstlisting}
func PushFrontiers (Partition PID)
 foreach (P in Partitions != PID) do
 @$|$@ NextFrontier[P] ==> Frontier[P]
 @$|$@         (local) ==> (remote)
 end for
end func
\end{lstlisting}
 \newline
 \newline
 \newline
\captionof{algorithm}{Pull Frontiers}
\label{alg:gather}
\begin{lstlisting}
func PullFrontiers (Partition PID)
 foreach (P in Partitions != PID) do
 @$|$@ Frontier[P] <== NextFrontier[P]
 @$|$@     (local) <== (remote)
 end for
end func
\end{lstlisting}
\end{minipage}
\end{minipage}
\subsection{Partition Specialization}%
\label{sec:partitioning}%
A performance-critical decision is graph partitioning; in particular, we need to identify the part of the graph that should be placed on the space-constrained GPUs such that overall performance is maximized. We first observe that even though the bottom-up steps can significantly improve performance for some BFS levels (due to the reduction in total edge checks), these bottom-up steps  take the longest out of the overall execution (Fig. \ref{fig:frontier}). Thus accelerating these steps is essential for overall processing performance, and our focus.

We partition such that the low-degree vertices are assigned to the GPUs. The intuition behind this decision is threefold: first, processing the many low-degree vertices in parallel fits the GPU compute model (i.e., many small computations with insignificant load imbalance); second, the low-degree vertices occupy a small amount of memory (as they are not attached to many edges), a critical issue to the space constrained GPUs; and third, and most importantly, processing the low-degree vertices during the bottom-up steps is the main bottleneck as we have empirically verified. As we argue in the next section, this partitioning solution adds one additional advantage: it makes it easier to decide when to switch to bottom-up processing without communicating between partitions.
\subsection{Switching Processing Direction for a Partitioned Graph}%
\label{sec:swapping}%
The direction-optimized algorithm requires coordinating all processing elements when the processing switches from top-down to bottom-up (after processing the first few BFS levels) then switching back to top-down processing. These decisions are generally taken based on global information \cite{beamer2011searching} \cite{you2014designing} (e.g., the anticipated size of the next frontier) yet obtaining a more precise estimate is costly on a platform that does not offer shared memory.

\textit{Top-down to bottom-up switch.} We estimate the next frontier based on a static percent of the edges out of the current frontier. This worked well in most executions on the scale-free graphs we have experimented with. However, when using this technique on our partitioned setup, it would normally be necessary to synchronize frontier information across each partition. However, as shown in Fig. \ref{fig:frontier}, the frontier is rapidly built by the few high degree vertices, while the low degree vertices have virtually no impact on the decision to switch as they are discovered later. For this reason, the coordinator for switching can be the partition responsible for the high degree vertices: the CPU. This method is less costly than communicating among partitions to precisely anticipate frontier size, while retaining nearly identical accuracy in predicting the optimal switch point.

\textit{Bottom-up to top-down switch.} The performance gains tends to be low from switching back to top-down processing as the the final BFS levels require little time anyways. For this reason, partitions return to top-down after a fixed number of steps, so that all partitions return at the same time without state communicating or voting.
\subsection{Optimizations to Improve Access Locality}%
\label{sec:reindexing}%
After partitioning, a vertex is identified by two elements: a global ID which corresponds to its place in the original graph, and a local ID, which corresponds to its place in the partition. This indexing provides flexibility that can be exploited as follows. First, since the partition retains the global ID, permutation of local IDs enables optimizations: we can reorder vertices in memory to improve local partition access locality \cite{sallinen2014exploring}. Second, the adjacency lists can be ordered in decreasing order of vertex connectivity, so that the highest degree vertex in the adjacency list comes first. This optimization shortens the bottom up-steps as the higher degree vertices are most likely to belong in the frontier, thus scanning the neighbour list has a higher chance to stop earlier (also noted by \cite{yasui2014fast}).


Finally, note that the optimizations discussed in this section are applicable to both CPU-only and GPU-accelerated setups. Indeed, such optimizations makes it even more challenging to show the benefits of a heterogeneous platform as they significantly improve the performance of the CPU-only baseline, and hence they expose the communication and coordination overheads. However, as we show in the next section, our optimizations related to reducing communication overheads successfully eliminate it as a potential bottleneck. 
\section{Experimental Results}%
\label{sec:experiments}%
\textbf{Software Platform.} \textsc{Totem} \cite{gharaibeh2014efficient}, the framework we use to support our exploration, hides the complexity of developing graph algorithms from the programmer by providing abstractions for communication, the ability to specify graph partitioning strategies, as well as common optimizations such as bitmap frontier representations and vertex and adjacency list ordering. We implemented the direction-optimized BFS algorithm on top of \textsc{Totem}, as well as the optimizations discussed -- it is important to note that both the GPU-accelerated and the CPU-only experiments use the same CPU kernel (i.e., they both apply the optimizations discussed in Section \ref{sec:reindexing}).

\noindent\textbf{Hardware Platform.} The experiments were executed on a single machine with two Intel Xeon E5-2670v2 processors with 10 cores at 2.5GHz and 512 GB of shared memory. The machine hosts two NVidia Kepler K40s with 2880 cores at 0.75GHz and 12 GB of memory each. The peak memory bandwidth of the host is 59.7GB/s, while on the GPU is 288GB/s.

\noindent\textbf{Methodology.} We employ the experimental methodology defined by Graph500 and GreenGraph500. These require computing the BFS parent of each vertex (as opposed to only its level). While \textsc{Totem} uses the CSR format and represents each undirected edge as two directed edges, we do report performance in \textit{undirected traversed edges per second (TEPS)}, as required by Graph500. Reported results are harmonic means over 100 executions. We measure power at the outlet using a WattsUP meter that samples at 1Hz. To get representative energy consumption, we run each experiment for 10 minutes (e.g., repeating searches).

\noindent\textbf{Workloads.} Unless otherwise mentioned, the synthetic graphs used are Scale30 [1B V, 16B E], built with the Graph500 reference code generator and parameters. The real-world graphs used are undirected versions of Twitter \cite{cha2010measuring} [52M V, 1.9B E], Wikipedia \cite{KONECT} [27M V, 601M E], and LiveJournal \cite{snapnets} [4M V, 69M E].
\subsection{The Impact of Specialized Partitioning}%
\begin{figure}%
	\centering%
	\begin{subfigure}{.5\textwidth}%
		\centering%
		\includegraphics[width=\linewidth,trim=3 2 3 3,clip]{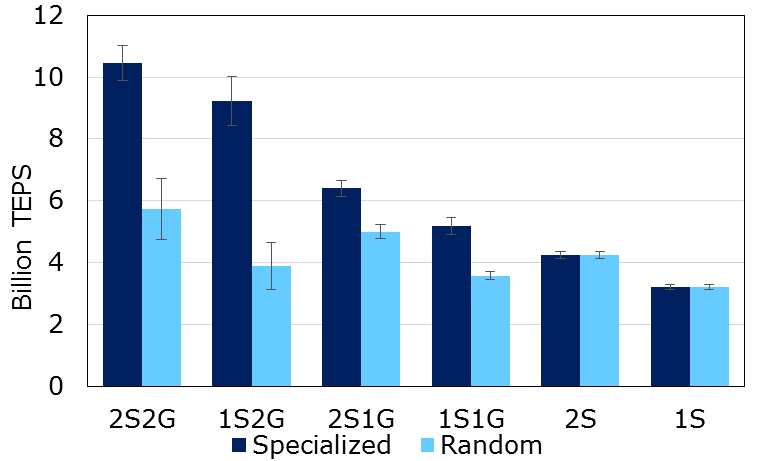}%
	\end{subfigure}%
    \begin{subfigure}{.5\textwidth}%
		\centering%
		\includegraphics[width=\linewidth,trim=3 2 3 3,clip]{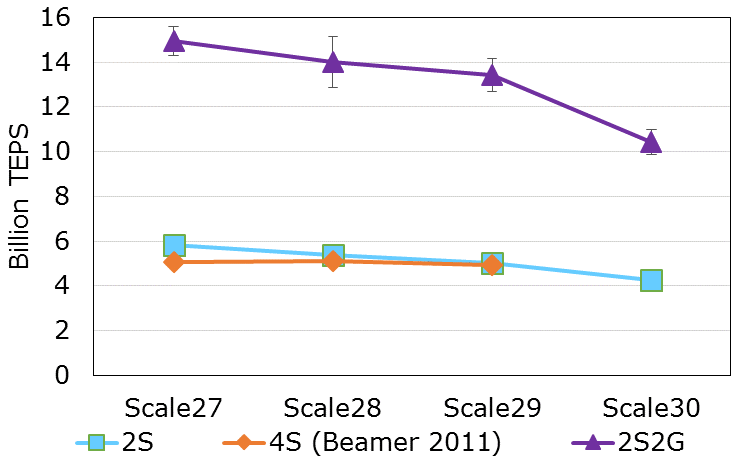}%
	\end{subfigure}%
	\caption{\textit{Left}: Direction-optimized BFS processing rate for specialized and random partitioning on hardware configurations with variable number of CPU Sockets (S) and and GPUs (G) for a Scale30 graph. \textit{Right}: Processing rates for synthetic graphs with size varying over one order of magnitude: Scale27 to Scale30. The curve labeled 4S presents the performance by Beamer \cite{beamer2011searching} on 4-Socket machine.}%
\label{fig:teps}%
\end{figure}

\label{sec:hardwareconf}%
Figure \ref{fig:teps} (\textit{left}) presents the processing rate for a Scale30 graph for configurations with one or two CPUs, and one or two GPUs. There are two takeaways: first, GPUs provide acceleration in all cases; and, relevantly for budget/energy-limited platforms, it is more efficient to add an additional GPU than an additional CPU. 

Second, and most importantly, the plot highlights the benefits of workload specialization: with random partitioning adding GPUs only offers a speedup proportional with memory footprint of the offloaded partition. \textit{The intelligent partitioning scheme we introduce offers a super-linear speedup: despite offloading only 8\% of the graph, 2 GPUs improve performance by 2.4x.}

Figure \ref{fig:teps} (\textit{right}) evaluates performance over multiple Graph500 sizes and shows that the GPU-accelerated setup with workload specialization consistently offers large gains. Larger-scale graphs tend to have a smaller TEPS rate due to lower data locality. We note that, despite the ability to allocate a larger part of the smaller graphs to the GPUs the gains level off for scale-free graphs: allocating more low-degree vertices becomes exponentially costly as the vertices have higher connectivity. The largest graph offers more potential for improvement if GPUs had more memory: 'only' 88\% of non-singleton vertices are allocated to the GPUs. This increases to 97\% for Scale29, and 99\% for Scale28; at which point there is not much room for performance gains from GPUs with larger memory.
\begin{figure}%
\begin{center}%
		\vspace{0cm}%
		\includegraphics[width=0.75\linewidth,trim=2 1 1 2,clip]{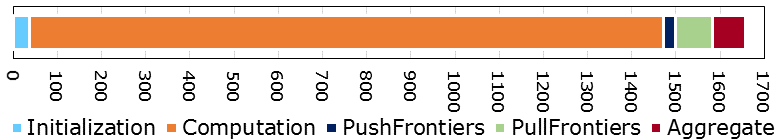}%
		\caption{BFS run time (ms) for a Scale30 graph broken down into components: initializing BFS status data, computation, and push- and pull-communication.}%
		\label{fig:total}%
\end{center}%
\end{figure}%

\textbf{Analyzing the Overheads.} Figure \ref{fig:total} highlights that performance is dominated by computation time: initialization, aggregation, and communication (presented separately for push/pull) are only a small fraction of the total runtime.

\begin{figure}%
	\centering%
    \begin{subfigure}{.5\textwidth}%
    	\centering%
        \includegraphics[width=\linewidth,trim=3 3 3 3,clip]{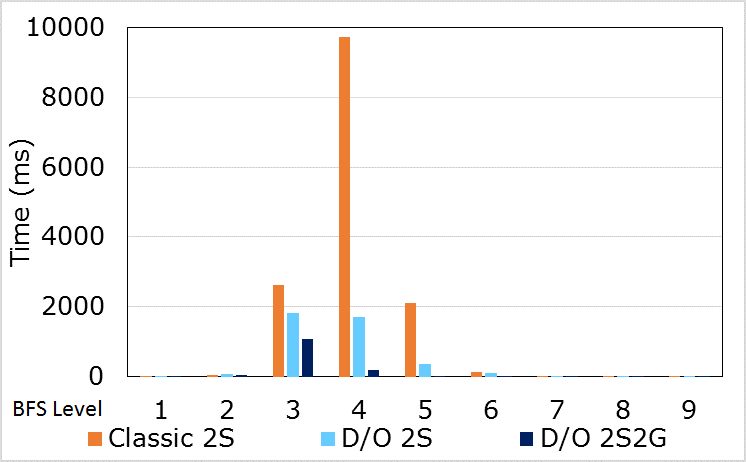}%
    \end{subfigure}%
    \begin{subfigure}{.5\textwidth}%
        \centering%
        \includegraphics[width=\linewidth,trim=3 3 3 3,clip]{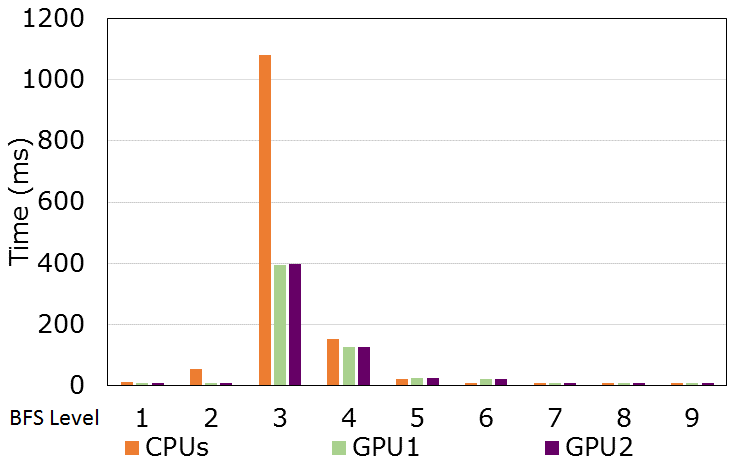}%
    \end{subfigure}%
    \caption{\textit{Left}: Per-level runtime (ms) for top-down (classic) and direction-optimized (D/O) BFS for a 2 CPU platform (2S), versus 2 CPUs and 2 GPUs (2S2G). \textit{Right}: Per-level execution time for CPUs/GPUs of the 2S2G platform for the direction-optimized execution in the left plot. \textit{Workload}: Scale30 graph.\\\\}%
	\label{fig:level}%
\end{figure}

Figure \ref{fig:level} (\textit{left}) breaks down the total runtime by BFS level for classic top-down BFS and direction-optimized BFS on a traditional (two CPU sockets -- labeled 2S) and hybrid (two CPUs and two GPUs -- labeled 2S2G) platform. The plot highlights two key points. First, it confirms the benefits of direction-optimization and it shows that these benefits are concentrated on faster processing of bottom-up levels 4 and 5. Second, it highlights the further gains offered by the hybrid platform, and pinpoints the gains to much faster level 4 processing.

As a result of the BSP model, the computation time is determined by the bottleneck processor in each step. Figure \ref{fig:level} (\textit{right}) presents the processing time per-level for each processing element: although occasionally (for levels 5 and beyond) the bottleneck is with the GPUs, the computation time for the initial bottom-up level (level 3) by the CPU dwarfs the rest of the execution time, leaving the other load-balancing inefficiencies nearly irrelevant.

\subsection{Comparison with Past Work using Real-world Graphs}%
\label{sec:realworld}%
We use real-world graphs to compare performance to that of the state-of-the-art graph processing framework \textsc{Galois}, whose direction-optimized BFS implementation compares favorably \cite{nguyen2013lightweight} to that of Ligra \cite{shun2013ligra}, PowerGraph \cite{gonzalez2012powergraph}, and GraphLab \cite{low2014graphlab}. (We run \textsc{Galois} on our experimental machine. We had extensive discussions with \textsc{Galois} authors to make sure comparisons are fair.) 

Table \ref{table:real} shows the following: first, as in Fig. \ref{fig:level} (\textit{left}) direction-optimized BFS largely outperforms top-down BFS. Second, our CPU-only versions of top-down and direction-optimized BFS perform largely similar to their \textsc{Galois} counterpart: this provides evidence that the baselines we used earlier for comparison to showcase the gains offered by our solution are fair. Furthermore, since in our hybrid algorithm the CPU and GPU kernels are executing concurrently, and the CPU is the bottleneck processor, improving our CPU kernel improves our overall execution rate, thus we have made all efforts to have efficient CPU-only kernels.

The hybrid direction-optimized version provides a performance boost of 2.0x for Twitter compared to the best CPU-only version. The larger diameter and less scale-free nature of the last two graphs reduce the impact of the direction-optimized approach, and expose more of the hybrid implementation overheads. Additionally, these smaller graphs expose less opportunity for the massive parallelism GPUs could harness. Nevertheless the hybrid implementation still offers a 1.3x speedup for LiveJournal and Wikipedia.

The table also highlights that the hybridization and the algorithm-level optimizations are synergetic, and together, they offer a significant boost in performance over generic and even optimized CPU versions. These results suggest that other scale-free real-world graphs will benefit from the techniques we propose.
\begin{table}%
\centering%
\caption{\textsc{Totem} and \textsc{Galois} (v2.2.1) performance in billion TEPS (higher is better), across real-world graphs. \textsc{Totem} executions use the same CPU kernel. The Naive kernel shown doesn't apply optimizations discussed in Section \ref{sec:reindexing}.}%
\label{table:real}%
\begin{tabular}{@{}llccccc@{}}%
\toprule%
                             & Algorithm           &\ Naive-2S\ \ &\ \textsc{Galois}-2S\ \ &\ \textsc{Totem}-2S\ \ & \textsc{Totem}-2S2G \\ \midrule
\multirow{2}{*}{Twitter}     & Top-Down            & 0.23     & 0.50      & 1.39     & 2.05       \\
                             & Direction-Optimized &          & 1.96      & 2.84     & \textbf{5.78}       \\ \midrule
\multirow{2}{*}{Wikipedia}   & Top-Down            & 0.84     & 0.42      & 1.14     & 1.29       \\
                             & Direction-Optimized &          & 1.12      & 1.49     & \textbf{2.01}       \\ \midrule
\multirow{2}{*}{LiveJournal} & Top-Down            & 0.54     & 0.99      & 1.26     & 1.57       \\
                             & Direction-Optimized &          & 1.23      & 1.96     & \textbf{2.59}       \\ \bottomrule
\\\\                            
\end{tabular}
\end{table}

\subsection{The Energy Case}%
\label{sec:energy}%
For Scale30 graphs, at 10.86 MTEPS/W, our CPU only implementation respectably ranked \#10 in the November 2014 Big Data category of the GreenGraph500 list\cite{GreenGraph500}. The hybrid configuration achieved over 2x better energy efficiency, ranking \#6 with 22.36 MTEPS/W. (We note that, our hybrid configuration ranked behind 5 similar submissions by the GraphCrest group \cite{yasui2014fast}, that all use more energy-efficient hardware: more and newer CPUs). 
For Scale29 graphs, on a recently acquired platform (2x Intel E5-2695, DDR4 memory, same GPUs) with 17.3GTEPS and 30.1 MTEPS/Watt we would rank at the top of today's Graph500 and, respectively, GreenGraph500.

The reason behind the energy gains the hybrid platform offers is that the GPU enables faster race-to-idle for the whole system (including energy expensive RAM), which means that the system draws high power for a significantly shorter period. Moreover, the most important factor that contributes to the energy gains is that the GPU, the processor with the higher Thermal Design Power (TDP), races-to-idle much faster than the CPUs (as shown in Fig. \ref{fig:level}). Finally we note that the property we observed for performance holds for energy efficiency: it is always better to add a GPU than a second CPU. For example, if we extrapolate the linear performance improvement from 1 CPU to 2 CPUs as in Fig. \ref{fig:teps} to 4 homogeneous CPUs, \textit{and conservatively assume these two new CPUs have no additional energy cost}, a system consisting of 4 of our CPUs would be approximately 16 MTEPS/W, still less efficient than our 2 CPU 2 GPU system.
\section{Summary}%
\label{sec:related}%
%
This work presents  the design, implementation and evaluation of a state-of-the-art BFS algorithm (Beamer et. al.'s direction-optimized algorithm \cite{beamer2011searching}) on top of a hybrid, GPU-accelerated platform. We present a number of critical optimizations that take advantage of both the characteristics of the hardware platform we target and common properties of many real-world datasets. We show that while the GPU has limited memory space, large-scale graphs can still benefit from GPU acceleration by carefully partitioning the graph such that the GPU is assigned the part of the workload that otherwise critically limits the overall performance. Moreover, we show that by applying simple yet effective optimizations, such gains are achieved even for discrete GPUs connected to the system via high-latency PCI bus. This offers a strong indication that these gains will hold for high-speed GPU platforms, such as AMD Fusion or NVLink.

Making progress on techniques able to harness heterogeneous platforms is essential in the context of current hardware trends: as the cost of energy continues to increase relative to the cost of silicon, future systems will host a wealth of different processing units. In this context, partitioning the workload and assigning the partitions to the processing element where they can be executed most efficiently in terms of power or time becomes a key issue.

\textbf{Acknowledgement.} This work was supported in part by the Institute for Computing, Information and Cognitive Systems (ICICS) at UBC.
\bibliographystyle{splncs03}
\bibliography{refs}
\end{document}